%% file: MN-orb_Kiefer.tex
\documentclass[useAMS,usenatbib]{mn2e}

\DeclareTextSymbol{\degre}{OT1}{23}
\usepackage{graphicx}
\usepackage{amsmath}
\usepackage{multirow}
\usepackage{afterpage}
\usepackage{pdflscape}
\usepackage{widetable}
\usepackage{tabularx}
\usepackage{placeins}

\title[Accurate SB2 orbits]{Masses of the components of SB2 binaries observed with {\it Gaia}. III. 
Accurate SB2 orbits for 10 binaries and masses of HIP\,87895 
\footnotemark[1]\thanks{based on observations performed
at the Observatoire de Haute--Provence (CNRS), France}}
\author[F. Kiefer et al.]{F. Kiefer$^{1}$\thanks{E-mail:
flavien@wise.tau.ac.il}, 
J.-L. Halbwachs$^{2}$, F. Arenou$^{3}$, D. Pourbaix$^{5}$, B. Famaey$^{2}$, P. Guillout$^{2}$,
  \newauthor
Y. Lebreton$^{3,4}$, A. Nebot G\'omez-Mor\'an$^{2}$, T. Mazeh$^{1}$, J.-B. Salomon$^{2}$, C. Soubiran$^{6}$ 
\newauthor
and L. Tal-Or$^{1,7}$\\
$^{1}$School of Physics and Astronomy, Tel Aviv University, Tel Aviv 69978, Israel\\
$^{2}$Observatoire astronomique de Strasbourg, Universit\'e de Strasbourg, CNRS, UMR 7550,
11 rue de l'Universit\'{e}, 67000 Strasbourg, France\\
$^{3}$GEPI, Observatoire de Paris, PSL Research University, CNRS, Universit\'e Paris Diderot, Sorbonne Paris Cit\'e, Place Jules Janssen, \\92195
Meudon, France\\
$^{4}$Institut de Physique de Rennes, Universit\'e de Rennes 1, CNRS UMR 6251, F-35042 Rennes, France\\
$^{5}$FNRS, Institut d'Astronomie et d'Astrophysique, Universit\'{e} Libre de Bruxelles, boulevard du Triomphe, 1050 Bruxelles, Belgium\\
$^{6}$Laboratoire d'Astrophysique de Bordeaux, LAB UMR 5804, CNRS, 2 rue de l'Observatoire, BP 89 33270 Floirac, France\\
$^{7}$Institut f\"ur Astrophysik (IAG), Friedrich-Hund-Platz 1, D-37077 G\"ottingen, Germany
}

\begin{document}

\date{Accepted . Received 2015 ; in original form 2015}

\pagerange{\pageref{firstpage}--\pageref{lastpage}} \pubyear{2015}

\maketitle

\label{firstpage}

\begin{abstract}
In anticipation of the {\it Gaia} astrometric mission, a large sample of spectroscopic binaries has been observed since 2010 with the SOPHIE spectrograph at the Haute--Provence Observatory. Our aim is to derive the orbital elements of double-lined spectroscopic binaries (SB2s) with an accuracy sufficient to finally obtain the masses of the components with relative errors as small as 1~\% when the astrometric measurements of {\it Gaia} are taken into account.
In this paper we present the results from five years of observations of 10 SB2 systems with periods ranging from 37 to 881 days. Using the {\sc todmor} algorithm we computed radial velocities from the spectra, and then derived the orbital elements of these binary systems. The minimum masses of the components are then obtained with an accuracy better than 1.2~\% for the ten binaries. Combining the radial velocities with existing interferometric measurements, we derived the masses of the primary and secondary components of HIP 87895 with an accuracy of 0.98\% and 1.2\% respectively. 
\end{abstract}



\section{Introduction}

Mass is the most crucial input in stellar internal structure modelling. It predominantly influences the luminosity of a star at a given stage of its evolution, and also its lifetime. The knowledge of the mass of stars in a non-interacting binary system, together with the assumption that the components have same age and initial chemical composition,  allows the age and the initial helium content of the system to be determined and therefore to characterize the structure and evolutionary stage of the components. Such modelling provides insights into the physical processes governing the structure of the stars. Moreover provided masses are known with great accuracy~\citep{Lebreton2005}, it gives constraints on the free physical parameters of the models. Therefore, modelling stars with extremely accurate masses (at the 1 \% level), in different ranges of masses, would allow to firmly anchor the models of the more loosely constrained single stars.

 This paper is the third in a series dedicated to the derivation of accurate masses of the components of double-lined spectroscopic binaries (SB2) with the forthcoming astrometric measurements from the {\it Gaia} satellite. In paper I \citep{Halb2014}, we have presented our program to derive accurate masses from {\it Gaia} and from high-precision spectroscopic observations. 
We have selected a sample of 68 SB2s for which we expect to derive very precise inclination with {\it Gaia}, and $M\sin^3 i$ with the Spectrographe pour l'Observation des PH\'enom\`enes des Int\'erieurs Stellaires et des Exoplan\`etes (SOPHIE spectrograph, Haute-Provence Observatory). Our objective is to determine for these SB2 systems the masses of the two components with an accuracy about 1~\%. 
We have been observing these stars since 2010 with SOPHIE. A first result of our program was the detection of the secondary component in the spectra of 20 binaries which were previously known as single-lined (paper\,I). A second result was the determination of masses for 2 SB2 with accuracy between 0.26 and 2.4\,\%, using astrometric measurements from PIONIER and radial velocities from SOPHIE (paper\,II).

Here, we present the accurate orbits measured for 10 SB2s (Table~\ref{tab:obs}) with periods ranging from 37 to 881 days. After 5 years of observations with SOPHIE, we collected a total of 123 spectra. In addition, a large number of previously published measurements is available for each of them in the SB9 catalog~\citep{SB9}. Four of these targets are new SB2 identified in paper I, and previously known as SB1. Finally, we combined the radial velocity (RV) measurements of one star (HIP~87895) with existing interferometric measurements and derive the masses of the two components.

The observations are presented in Section~\ref{sect:observations}. The method of measurements of radial velocities from SOPHIE's observations is explained in Section~\ref{sect:RV}. We derive the orbital solutions in Section~\ref{sec:orbits}, discussing in particular the issue of the uncertainties when combining different datasets from different instruments. The derivation of the masses of HIP\,87895 is discussed in Section~\ref{sec:HIP87895}. Finally, we summarize and conclude on our findings in Section~\ref{sec:conclusion}.

\section{Observations}
\label{sect:observations}

The observations were performed at the T193 telescope of the Haute-Provence Observatory, with the SOPHIE spectrograph. SOPHIE is dedicated to the search of extrasolar planets, and, thanks to its high resolution ($R$$\sim $$75,000$), it enables accurate stellar radial velocities to be measured for SB2 components.

 The spectra were all reduced through SOPHIE's pipeline, including localization of the orders on the frame, optimal order extraction, cosmic-ray rejection, wavelength calibration, flat-fielding and bias subtraction. The minimum signal-to-noise ratio (SNR) is 40 for the faintest stars of the sample, but it may be as large as 150 for a 6-magnitude star. Before each observation run ephemerides were derived from existing orbits provided by the SB9 catalogue \citep{SB9}, and priority classes were assigned on the basis of the orbital phase. Four classes were used: the lowest priority corresponds to stars
with expected RVs of primary and secondary component sufficiently different to permit accurate measurements, and the highest priority is reserved to the observations of the periastron of eccentric orbits.

Among all the observed SB2, we have selected those which were observed over at least one period, and which received a minimum of 11 observations. Table~\ref{tab:obs} summarizes this information. Given the very high quality of the measurements, an SB2 orbit could be derived in principle from only 6 of those observations, provided they were made at the most relevant phases. However, we show in Section~\ref{sec:orbits} that 11 observations are necessary to validate the RV uncertainties, and to correct them when necessary.

\begin{table}
\caption{The ten SB2 analyzed in this paper. }
\small
\begin{tabular}{r@{~~}c@{~~}c@{~~}c@{~~}c@{~~}c@{~~}c@{~~}c@{~~}l}
\hline
HIP     & HD    &  V     & Period$^a$   & $N_\text{spec}^{\phantom{spec}b}$& Time span$^c$    & SNR$^d$  \\
        &    & (mag.) &  (day)   &                 &   (period)   &      & \\
\hline   
\multicolumn{8}{c}{\it Previously published SB2} \\[1ex]
12081  &  15850   & 7.72   & 443.49  & 11              & 3.3          & 80    \\
17732  &  23626   & 6.27   & 277.89  & 11              & 5.5          & 120  \\
56275  &  100215   & 7.99   & 47.88   & 13              & 30.7         & 80   \\
87895  &  163840   & 6.33   & 880.78  & 14              & 2.1          & 120  \\
95575  &  183255   & 8.05   & 166.36  & 13              & 11.9         & 75   \\
100321 &  195850   & 7.02   & 37.94   & 11              & 45.3         & 80   \\
\hline
\multicolumn{8}{c}{\it SB2 identified in paper I, previously published as SB1} \\[1ex]
13791  &  18328    & 8.87   & 48.71   & 13              & 28.8         & 40    \\
61727  &  110025   & 7.58   & 54.88   & 12              & 28.3         & 80    \\
62935  &  120005   & 8.53   & 139.00  & 11              & 10.6         & 40    \\
67195  &  120005   & 6.50   & 39.28   & 14              & 37.3         & 120   \\
\hline
\end{tabular}\\
\flushleft 
$^a$ The period values are taken from the SB9 catalog \citep{SB9}. \\
$^b$ $N_\text{spec}$ gives the number of spectra collected with SOPHIE. \\
$^c$ The time span is the total span of observation epochs, counted in number of periods. \\
$^d$ SNR is the median signal-to-noise ratio of each sample.
\label{tab:obs}
\end{table}

\section{Radial velocity measurements}
\label{sect:RV}

The radial velocities of the components are derived using the TwO-Dimensional CORrelation algorithm {\sc todcor} \citep{zucker94,zucker04}.
It calculates the cross-correlation of an SB2 spectrum and two best-matching stellar atmosphere models, one for each component of the observed binary system. This two-dimensional cross-correlation function (2D-CCF) is maximized at the radial velocities of both components. The multi-order version of {\sc todcor}, named {\sc todmor}~\citep{zucker04}, determines the radial velocities of both components from the gathering of 2D-CCF obtained from each order of the spectrum. 

All SOPHIE multi-orders spectra were corrected for the blaze using the response function provided by SOPHIE's pipeline; then for each of them, the pseudo-continuum was detrended using a p-percentile filter~(paper II, or e.g. \cite{Hodg1985}).

For both components of each binary, we determined a best-matching theoretical spectra from the PHOENIX stellar atmosphere models~\citep{Huss2013}. We optimized the 2D-CCF with respect to effective temperature $T_\text{eff}$, rotational broadening $v \sin i$, metallicity [Fe/H], surface gravity $\log(g)$, and flux ratio at $4916$\,\AA, $\alpha$$=$$F_2/F_1$. Furthermore, each theoretical spectrum is convolved with the instrument line spread function, here modeled by a Gaussian, and pseudo-continuum detrended like the observed spectrum. The spectral parameters obtained through this method are given in Table~\ref{tab:stellpar}. We determined spectral parameters from 4 spectra per binary on average, with the two components well individualized. The values and their uncertainties given in Table~\ref{tab:stellpar} are the average and standard deviation of the individual estimations. The 1$\sigma$ uncertainties do not include known systematics of theoretical models with respect to real spectral types (see e.g.~\cite{Torres12}).  

It is worth mentioning that the derived metallicity [Fe/H] is systematically subsolar. Optimizing the CCF of several spectra of the Sun obtained by observing Vesta and Ceres with SOPHIE gave spectral parameters consistent with the known values for the Sun except metallicity that was found to be -0.33 dex. However, we kept the values of metallicity that maximized the 2D-CCF, as given in Table~\ref{tab:stellpar}.

\begin{table}
\begin{minipage}{\columnwidth}
\caption{\label{tab:stellpar}The stellar parameters determined by optimization of the 2D-CCF obtained with {\sc todmor}. Explanations in Section~\ref{sect:RV}.}
\small
\begin{tabular}{@{}lrrrrr@{}}
\hline 
HIP     & $ T_\text{eff,1}$ &  $ \log g_1$ &  $ V_1 \sin i_1$ $^a$  &  $[\text{Fe/H}]$ &  $\alpha$   \\ 
HD      & $ T_\text{eff,2}$ &  $ \log g_2$ &  $ V_2 \sin i_2$ $^a$ &                 &                \\ 
        & (K)               &  (dex)       &   (km s$^{-1}$)   &     (dex)       & (flux ratio)   \\ 
\hline
12081 & 6290       & 4.24        & 11.9       & -0.37       & 0.635 \\ 
15850 & $\pm$ 23   & $\pm$0.05   & $\pm$0.1   & $\pm$0.01  & $\pm$0.012  \\ 
      & 6003       & 4.30        & 6.5        &            &        \\ 
      & $\pm$ 13   & $\pm$0.01   & $\pm$0.1   &            &           \\ [1ex]
13791 & 6173       & 4.31        & 5.5       & -0.34       & 0.034  \\   
18328 & $\pm$ 6    & $\pm$0.01   & $\pm$0.1  & $\pm$0.01   & $\pm$0.001  \\ 
      & 4953       & 5.09        & 0         &             &        \\   
      & $\pm$ 95   & $\pm$0.01   &           &             &              \\[1ex] 
17732 & 6030       & 3.34        & 9.4       & -0.71       & 0.258  \\    
23626 & $\pm$  7   & $\pm$0.03   & $\pm$0.1  & $\pm$0.01   & $\pm$0.004  \\ 
      & 6051       & 3.84        & 0         &             &        \\    
      & $\pm$ 18   & $\pm$0.01   &           &             &        \\ [1ex]
56275 & 6889       & 3.93        & 20.6       & -0.33       & 0.036 \\   
100215& $\pm$  36  & $\pm$0.11   & $\pm$1     & $\pm$0.01   & $\pm$0.002  \\ 
      & 4906       & 4.33        & 0          &             &        \\    
      & $\pm$ 75   & $\pm$0.01   &            &             &              \\ [1ex]
61727 & 6461       & 3.82        & 9.8       & -0.48       & 0.070  \\    
110025& $\pm$  31  & $\pm$0.15   & $\pm$0.1  & $\pm$0.06 & $\pm$0.007  \\ 
      & 5256       & 4.31        & 7.7       &             &          \\  
      & $\pm$ 168  & $\pm$0.14   & 1.4       &             &              \\ [1ex]
62935 & 5818       & 4.03        & 4.9       & -0.27        & 0.054  \\   
112138 & $\pm$ 85  & $\pm$0.08   & $\pm$0.4  & $\pm$0.01   & $\pm$0.015  \\ 
      & 4378       & 4.40        & 0         &             &        \\  
      & $\pm$ 115  & $\pm$0.13   &           &             &           \\ [1ex]
67195 & 6411       & 4.29        & 13.6       & -0.21        & 0.027 \\  
120005& $\pm$ 29   & $\pm$0.13   & $\pm$0.2   & $\pm$0.03   & $\pm$0.002  \\ 
      & 4478       & 4.72        & 4.5        &             &        \\   
      & $\pm$ 418  & $\pm$0.21   & $\pm$0.1   &             &           \\ [1ex]
87895 & 5970       & 4.33        & 4.7        &  -0.19       & 0.036  \\   
163840& $\pm$  1  & $\pm$0.01   & $\pm$0.1    & $\pm$0.01 & $\pm$0.004  \\ 
      & 4385       & 4.81        & 0          &             &          \\ 
      & $\pm$ 134  & $\pm$0.04   &            &             &               \\ [1ex]
95575 & 4908       & 4.75        & 3.4       & -0.88       & 0.431  \\   
183255& $\pm$ 5    & $\pm$0.03   & $\pm$0.1  & $\pm$0.01 & $\pm$0.008  \\ 
      & 4088       & 4.51        & 0         &             &          \\  
      & $\pm$ 5    & $\pm$0.03   &           &             &               \\ [1ex]
100321 & 6485       & 4.17       & 14.2       & -0.39       & 0.207 \\   
195850 & $\pm$  6   & $\pm$0.05  & $\pm$0.2   & $\pm$0.01   & $\pm$0.011 \\ 
       & 5558       & 4.39       & 5.5        &             &        \\    
       & $\pm$ 62   & $\pm$0.04  & $\pm$ 1    &             &           \\ 
\hline
\end{tabular}
\end{minipage}
$^a$ A null value is given to $V\sin i$ with no error bar whenever found less than SOPHIE's typical pixel size $\sim$2km\,s$^{-1}$.\\

\end{table}

We then applied {\sc todmor} to all multi-order spectra of each target and determined the radial velocities of both components discarding all orders harboring strong telluric lines. For each of the non-discarded orders, we calculated a two dimensional cross-correlation function, and used the maximum of this function to derive radial velocities for the primary and the secondary. 
Final velocities for each component are obtained by averaging these measurements and incorporating a correction for order-to-order systematics -- typically 200-500\,m\,s$^{-1}$. The final velocities are displayed in Table~\ref{tab:RVs}. They are used to derive the orbital solutions for the 10 SB2 in the next section.

\begin{table*} 
\caption{\label{tab:RVs} New radial velocities from SOPHIE and obtained with {\sc todmor}. Outliers are marked with an asterisk ($^*$) and are not taken into account in the analysis.}
\scriptsize
\begin{minipage}{89mm}
\input{HIP12081_RV_table_latex}
\end{minipage}%
\begin{minipage}{89mm}
\input{HIP13791_RV_table_latex}
\end{minipage}\\%
\begin{minipage}{89mm}
\input{HIP17732_RV_table_latex}
\end{minipage}%
\begin{minipage}{89mm}
\input{HIP56275_RV_table_latex}
\end{minipage}\\%
\begin{minipage}{89mm}
\input{HIP61727_RV_table_latex}
\end{minipage}%
\begin{minipage}{89mm}
\input{HIP62935_RV_table_latex}
\end{minipage}\\
\begin{minipage}{89mm}
\input{HIP67195_RV_table_latex}
\end{minipage}%
\begin{minipage}{89mm}
\input{HIP87895_RV_table_latex}
\end{minipage}
\vspace*{1cm}
\end{table*}

\addtocounter{table}{-1}
\begin{table*}
\caption{Continued.}
\scriptsize
\begin{minipage}{89mm}
\input{HIP95575_RV_table_latex}
\end{minipage}%
\begin{minipage}{89mm}
\input{HIP100321_RV_table_latex}
\end{minipage}
\vspace*{1cm}
\end{table*}


\section{Derivation of the orbits}
\label{sec:orbits}
 
The orbital solutions for the 10 SB2 are derived by combining the new measurements presented in this paper with previously published RVs (references in Table~\ref{tab:corsigRVprev}). 

Since several datasets coming from different instruments are used together to derive a common orbital solution, realistic errors should be attributed to each dataset properly. It is explained in the following section. This process guarantees that each dataset receives the proper weight with respect to all others, including the new SOPHIE measurements presented here. 

\subsection{Correction of uncertainties}
\label{sec:corer}

\begin{table*}
\caption{Correction terms applied to the uncertainties of the previous and of the new RV measurements. The composition of these terms into a uncertainty correction is explained in Section~\ref{sec:corer}, eqs.~\ref{eq:correction1} and~\ref{eq:correction2}. 
}
\begin{tabular}{rlcccccccc}
\hline
HIP    & Reference of previous RV  &  \multicolumn{4}{c}{Correction terms for previous measurements} &  \multicolumn{4}{c}{Correction terms for new measurements}   \\
       &                        &$\varepsilon_{1,p}$&$\varphi_{1,p}$&$\varepsilon_{2,p}$&$\varphi_{2,p}$&$\varepsilon_{1,n}$&$\varphi_{1,n}$& $\varepsilon_{2,n}$& $\varphi_{2,n}$\\
       &                        &km s$^{-1}$        &               & km s$^{-1}$       &               & km s$^{-1}$       &               &  km s$^{-1}$       &          \\
\hline
12081  & \cite{Griffin05}       &       0           & 0.615         & 0                 & 0.565         &  0.1428           & 1             & 0.0061             & 1        \\
13791  & \cite{Imbert06}        & 0.24              & 1             & $\cdot$           & $\cdot$       & 0.0287            &  0.885        & 0.2689             & 0.885 \\
17732  & \cite{GriffinSuchkov03}& 0                 & 0.272         & 0                 & 0.759         & 0                 &  1.101        & 0.0380             &  1.101  \\
56275  & \cite{Griffin06}       & 0                 & 1.208         & 0                 & 1.208         & 0.6909            &  0.905        & 0.2315             &  0.905\\
61727  & \cite{Halb12}          & 0.3212            & 1             & $\cdot$           & $\cdot$       & 0.0261            &  0.911        & 0.4914             &  0.911\\
62935  & \cite{Griffin04}       & 0                 & 0.594         & $\cdot$           & $\cdot$       & 0.0294            & 1.101         &  0.1774            & 1.101  \\
67195  & \cite{Shajn39}         & 0                 & 5.243         & $\cdot$           & $\cdot$       & 0.0175            & 1.046         & 0                  & 0.8170 \\
87895  & \cite{McAlister95}     & 0                 & 0.342         & 0                 & 0.342         & 0.0355            & 1.329         & 0.0662             &  1.329\\
95575  & \cite{Toko91}          & 1.022             & 1             & 1                 & 1             & 0.0108            & 0.897         & 0.0399             &  0.897\\
100321 & \cite{Carquillat00}    & 0                 & 0.541         & 0                 & 1.097         & 0.0665            & 1.017         & 0                  & 0.572 \\
\hline
\label{tab:corsigRVprev}
\end{tabular}
\vspace*{1cm}
\end{table*}

Uncertainties of previously published measurements, when provided, are usually underestimated. On the other hand, many lists of RV measurements do not include uncertainties, but only weights (W). Therefore, two different procedures are applied to attribute correct uncertainties to these retrieved measurements. They are both based on the calculation of the $F_2$ estimator of the goodness-of-fit \citep[see Paper\,II, equation 1, or][]{Kendall}:

\begin{itemize}
\item
When the uncertainties are provided, a noise is quadratically added to the original uncertainties, in order to get exactly $F_2$=$0$
for the SB1 orbit of each component. 
Since the original uncertainties are underestimated, this results in decreasing the variations of the
relative weights of the measurements of a given component. 

\item
When only weights are given, they are first converted to uncertainties ($\sigma=\sqrt{1/W}$). Then they are scaled in order to get $F_2$=$0$ for the SB1 orbit of each component.
\end{itemize}

After this transformation, the SB2 orbit is derived, and $F_2$ is considered again. The final uncertainties are obtained by multiplying the ones derived above with a factor chosen in order to get $F_2$=$0$. All these operations result in applying the following formulae:

\begin{align}
\sigma^\text{corr}_{RV, 1} &= \varphi_1 \times \sqrt{\sigma_{RV, 1}^2 + \varepsilon_1^2} \label{eq:correction1}\\
\sigma^\text{corr}_{RV, 2} &= \varphi_2 \times \sqrt{\sigma_{RV, 2}^2 + \varepsilon_2^2} \label{eq:correction2}
\end{align}

Table~\ref{tab:corsigRVprev} lists the derived values of the correction terms $\varphi_1$, $\varphi_2$, $\varepsilon_1$ and $\varepsilon_2$ for all stars of our sample.

The same procedure is applied to the uncertainties derived by {\sc todmor} for the new measurements. For two binaries (HIP\,67195 and HIP\,100321), the SB1 orbit of the secondary component leads to a negative value of $F_2$, implying that the uncertainties are slightly overestimated. In order to get $F_2$$=$$0$, we prefered to keep the relative weights fixed, and simply multiply the uncertainties by a coefficient $\varphi_{2}$ lower than 1.

All the RV measurements and the uncertainties used in the derivation of the orbits will be available through
the SB9 Catalogue \citep{SB9}, which is accessible on-line\footnote{http://sb9.astro.ulb.ac.be/}.

\subsection{Derivation of the orbital elements}
\label{sect:orbsol} 

For each binary, we fitted an SB2 orbital model to the previously published datasets combined with the new SOPHIE observations. The parameters were optimized using a Levenberg-Marquard method. 

The final orbital solution consists in the following orbital elements: the period, $P$, the eccentricity, $e$, the epoch of the periastron, $T_0$, the longitude of the periastron for the primary component, $\omega_1$, the RV semi-amplitudes of each component, $K_1$ and $K_2$, and the RV of the barycentre, $V_0$. 

We also added 3 supplementary parameters, which are: a systematic offset between the measurements previously published and the new ones, $d_{n-p}$, and the offsets between the RVs of the primary and of the secondary component, $d_{2-1}^p$ and $d_{2-1}^n$. 
The offset $d_{2-1}^p$ is usually due to the fact that the published RVs were obtained from templates which are not specifically adapted to each component. For instance, a spectrovelocimeter like CORAVEL \citep{Baranne1979} was working by projecting any spectrum on a mask representing the spectrum of Arcturus. 
The offset between the primary and secondary RVs derived with {\sc todmor}, $d_{2-1}^n$ is expected to be null, but it is significant for some stars, since the spectra of the PHOENIX library do not perfectly represent the actual ones.

\begin{table*}
\centering
\caption{The orbital elements derived from the previously published RV measurements and from the
new ones. The radial velocity of the barycentre, $V_0$, is in the reference system of the new measurements of the primary component. 
The minimum masses and minimum semi-major axes are derived from the true period 
($P_{true}=P \times (1-V_0/c)$).}
\scriptsize
\setlength{\tabcolsep}{2mm}
  \begin{tabular}{@{}lrrrrrrrrrrrrrr@{}}
  \hline
HIP & $P$ & $T_0$(BJD) & $e$ &  $V_0$ & $\omega_1$ & $K_1$ &  ${\cal M}_1 \sin^3 i$ &  $a_1 \sin i$&$N_1$& $d_{n-p}$ & $\sigma(O_1-C_1)_{p,n}$ \\
HD  &     &           &     &        &            & $K_2$ &  ${\cal M}_2 \sin^3 i$ &  $a_2 \sin i$&$N_2$& $d^p_{2-1},d^n_{2-1}$ &$\sigma(O_2-C_2)_{p,n}$    \\
    & (d) & 2400000+  &  &(km s$^{-1}$)&($^{\rm o}$)&(km s$^{-1}$)&(${\cal M}_\odot$)&  (Gm) & & (km s$^{-1}$) &(km s$^{-1}$)    \\
  \hline
  \\
12081  & 443.288     & 55905.74   & 0.58494      & -7.883     & 283.66    & 16.784    &0.5493     & 82.98     &47+11& -1.141        &0.634,0.128\\
15850  &$\pm 0.023$  &$\pm 0.12$  &$\pm 0.00054 $&$\pm  0.046$&$\pm 0.10$ &$\pm 0.057$&$\pm0.0019$&$\pm 0.28$ &     & $\pm 0.101$   &         \\
       &             &            &              &            &           & 18.258    &0.5050     &90.261     &47+11& -0.208, -0.064&0.548,0.020\\
       &             &            &              &            &           &$\pm 0.015$&$\pm0.0034$&$\pm 0.052$&     & $\pm 0.124$, $\pm0.050$ & \\
&&&&&&&&&&&\\ 
13791  & 48.70895    & 55050.392  & 0.18781      &-8.537       &55.26     &  16.030   &0.2405      &10.5456    &33+12&0.686          &0.431,0.022\\
18328  &$\pm 0.00023$&$\pm 0.043$ &$\pm 0.00078$&$\pm 0.012$  &$\pm 0.29$&$\pm 0.015$ &$\pm0.0027$ &$\pm0.0090$&     & $\pm 0.078$   &         \\
       &             &            &             &             &          & 27.07      &0.14241     &17.809     &0+11 & $\cdots$ , 0.400      &$\cdots$ ,0.246 \\
       &             &            &             &             &          &$\pm 0.14$  &$\pm0.00094$&$\pm 0.089$&     & $\cdots$ ,$\pm 0.096$ &        \\
&&&&&&&&&&&\\
17732  & 277.9500    & 55230.207  & 0.14319     &-2.918       &159.14    &  18.6644   &1.1359    &70.598    &41+11&-1.096         &0.505,0.0080\\
23626  &$\pm 0.0074$ &$\pm 0.095$ &$\pm 0.00083$&$\pm 0.011$ &$\pm 0.14$&$\pm 0.0051$&$\pm0.0019$&$\pm 0.022$&     & $\pm 0.058$   &         \\
       &             &            &             &             &          & 23.208     &0.9135    &87.784     &41+11& 0.279, 0.116 &1.670,0.054 \\
       &             &            &             &             &          &$\pm 0.017$ &$\pm0.0010$&$\pm 0.067$&    & $\pm 0.169,\pm 0.027$ &    \\
&&&&&&&&&&&\\
56275  & 47.89410    & 56781.809  & 0.2317      &-17.879      &247.52    &  28.00     &1.5007     &17.94      &31$^a$+13&-1.273         &1.299,0.571\\
100215 &$\pm 0.00089$&$\pm 0.080$ &$\pm 0.0018$ &$\pm 0.18$   &$\pm 0.71$&$\pm 0.18$  &$\pm0.0086$&$\pm 0.11$ &     & $\pm 0.294$   &         \\
       &             &            &             &             &          & 51.709     &0.8126     &33.128     &10+13& 0.963, 0.273 &3.786,0.207 \\
       &             &            &             &             &          &$\pm 0.087$ &$\pm0.0090$&$\pm 0.051$&    & $\pm 1.247,\pm 0.202$ &   \\
&&&&&&&&&&&\\
61727  & 54.87864    & 56476.6509 & 0.34744     &-1.5628      &185.679   &  30.842    &1.432      &21.823     &43+12& 0.168     &0.639,0.020\\
110025 &$\pm 0.00018$&$\pm 0.0072$&$\pm 0.00051$&$\pm 0.0093$ &$\pm 0.057$&$\pm 0.019$&$\pm0.017$ &$\pm 0.011$&     &$\pm 0.097$&         \\
       &             &            &             &             &          & 48.51      &0.9104     &34.33      &0+12 & $\cdots$ , 0.358  &$\cdots$ ,0.435 \\
       &             &            &             &             &          &$\pm 0.26$  &$\pm0.0060$&$\pm 0.18  $&    &$\cdots$ ,$\pm 0.142$&        \\
&&&&&&&&&&&\\
62935  &139.0081     & 55680.66   & 0.1508      &0.244        &31.37     &  15.643    &0.4791     &29.557     & 89+11 &-0.189       &0.680,0.022\\
112138 &$\pm 0.0027$ &$\pm 0.17$  &$\pm 0.0018$ &$\pm 0.011$  &$\pm 0.44$&$\pm 0.023$ &$\pm0.0049$&$\pm 0.038$&       &$\pm 0.065$  &         \\
       &             &            &             &             &          & 23.03      &0.3254     &43.51      &0+11   & $\cdots$, 0.416   & $\cdots$,0.245 \\
       &             &            &             &             &          &$\pm 0.11$  &$\pm0.0020$&$\pm 0.20$ &       & $\cdots$,$\pm 0.071$&        \\

&&&&&&&&&&&\\
67195  &39.284974    & 56211.2841 & 0.78584     & -9.532      &325.250   &  45.44    &1.1731      &15.181     & 37+14 &-1.113       &5.027,0.021\\
120005 &$\pm 0.000091$&$\pm 0.0045$&$\pm 0.00050$&$\pm 0.016$ &$\pm 0.031$&$\pm 0.12$&$\pm0.0062$ &$\pm 0.026$&       &$\pm 0.864$  &         \\
       &             &            &             &             &          & 78.86     &0.6760      &26.346     &0+14   & $\cdots$ , -0.013   &$\cdots$ ,0.118 \\
       &             &            &             &             &          &$\pm 0.22$ &$\pm0.0034$&$\pm 0.048$ &       &$\cdots$ ,$\pm 0.050$&        \\
&&&&&&&&&&&\\
87895  & 881.629   & 55650.40 & 0.4165      &-32.346    &135.47    &  11.393   &0.9869     &125.57    &106+14&0.456      &0.842,0.035\\
163840 &$\pm 0.065$&$\pm 0.49$&$\pm 0.0014$ &$\pm 0.015$&$\pm 0.26$&$\pm 0.021$&$\pm0.0097$&$\pm 0.22$&      &$\pm 0.047$&      \\
       &           &          &             &           &          & 17.374    &0.6471     &191.50    &16+13 &-0.754, 0.319 &1.210,0.210\\
       &           &          &             &           &          &$\pm 0.077$&$\pm0.0041$&$\pm0.84$ &      &$\pm0.608$,$\pm0.058$&\\
&&&&&&&&&&&\\
95575  & 166.8351   & 56087.993 & 0.13698     &-64.2928    & 63.21    &  14.0202   &0.23291     &31.866     &23+12&0.394       &1.225,0.0098\\
183255 &$\pm 0.0031$&$\pm 0.094$&$\pm 0.00031$&$\pm 0.0040$&$\pm 0.20$&$\pm 0.0048$&$\pm0.00051$&$\pm 0.011$&     &$\pm 0.233$ &         \\
       &            &           &             &            &          & 15.696     &0.20804     &35.675     &9+12 &0.356, 0.097&1.981,0.044\\
       &            &           &             &            &          &$\pm 0.016$ &$\pm0.00028$&$\pm 0.037$&     &$\pm 0.592$,$\pm 0.014$&\\
&&&&&&&&&&&\\
100321 & 37.939920   & 56254.291  & 0.18242     & 0.941       &111.043   &  34.306    &1.06199   &17.596     &52+11   &0.324       &0.530,0.068\\
195850 &$\pm 0.000081$&$\pm 0.010$ &$\pm 0.00025$&$\pm 0.023$   &$\pm 0.096$&$\pm 0.025$&$\pm0.00081$&$\pm 0.013$&      &$\pm 0.079$ &         \\
       &             &            &             &             &          & 45.0916    &0.8080    &23.1284      &49+11 &0.493, 0.121&1.160,0.011\\
       &             &            &             &             &          &$\pm 0.0089$&$\pm0.0011$&$\pm 0.0041$&      &$\pm 0.174,\;\pm 0.028$&\\
 \hline
\label{tab:orbSB2}
\end{tabular}
\flushleft
$^a$ The 3 first measurements in \cite{Griffin06} were discarded.
\end{table*}

\subsection{Results}
\label{sect:results}

Even with a correction, the uncertainties of the measurements from SOPHIE remain small and have weights much larger than the others. Therefore, incorporating the published datasets in the calculation essentially improved the accuracy of the period determined. 
Conversely, the significant precision of the new SOPHIE measurements allowed us to reach a very good accuracy on all other orbital parameters, especially the minimum masses. 

The derived orbital elements for the 10 SB2 are given in Table \ref{tab:orbSB2}. Among the 20 component stars, 15 of them have $M \sin^3 i$ determined with an accuracy better than 0.7~\%, while only 5 stars have minimum mass accuracies between 1 and 1.2\%. 

A comparison between the standard deviation of the residuals of the previous measurements and of the new ones also illustrates the
improvements due to SOPHIE. It results from the last column of Table \ref{tab:orbSB2} 
that $\sigma(O-C)_p$ is between 2.3 and 239 times larger than $\sigma(O-C)_n$, with a median around 30. 

\begin{figure*}
\includegraphics[clip=,height=10.5 cm,width=150mm]{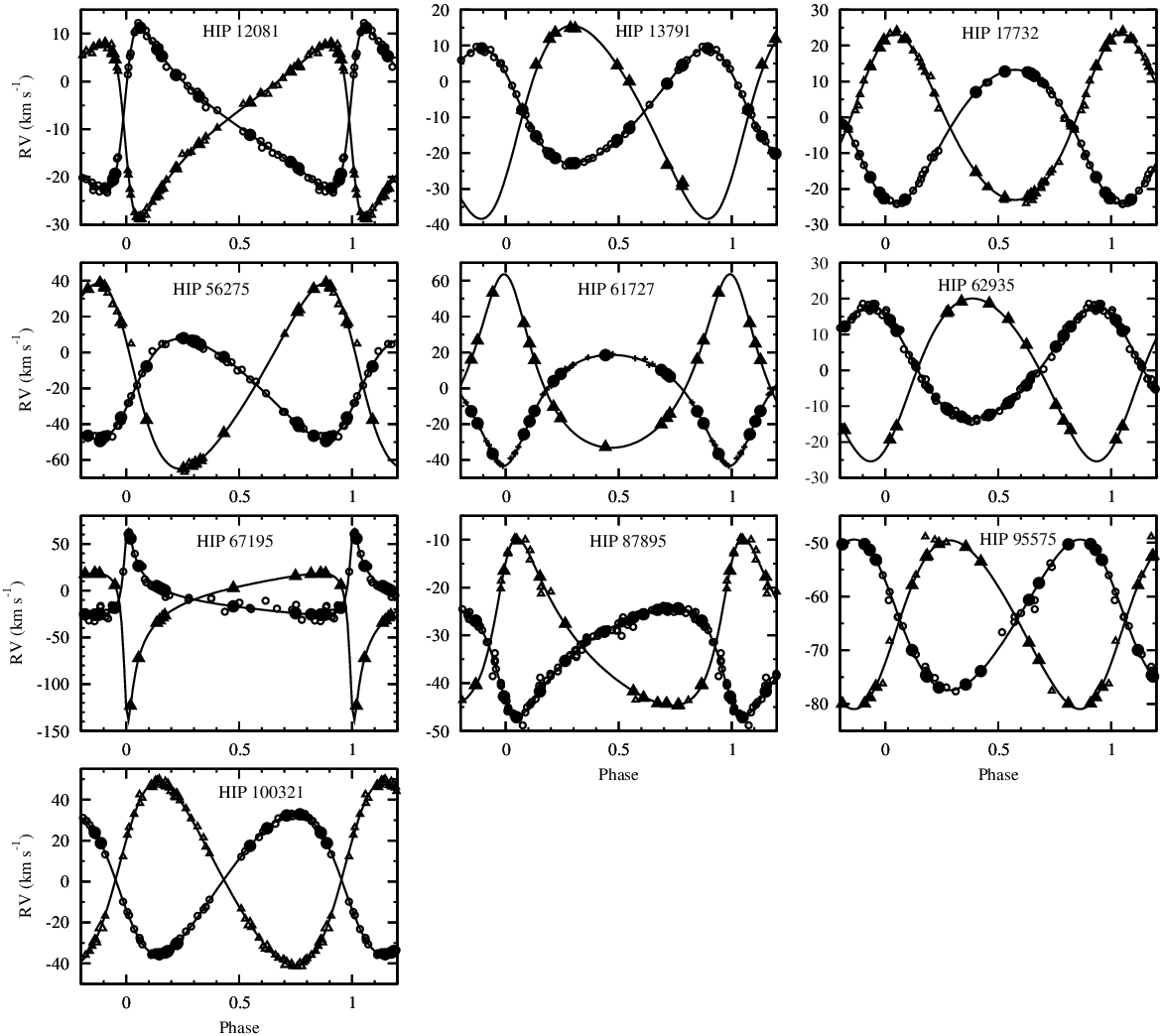} 
 \caption{The spectroscopic orbits of the 10 SB2; the circles refer to the primary component, and the
triangle to the secondary; the large filled symbols refer to the new RV measurements obtained with SOPHIE. For each SB2, the RVs are shifted to the
zero point of the SOPHIE measurements of the primary component.}
\label{fig:orbSB2}
%
\includegraphics[clip=,height=10.5 cm,width=150mm]{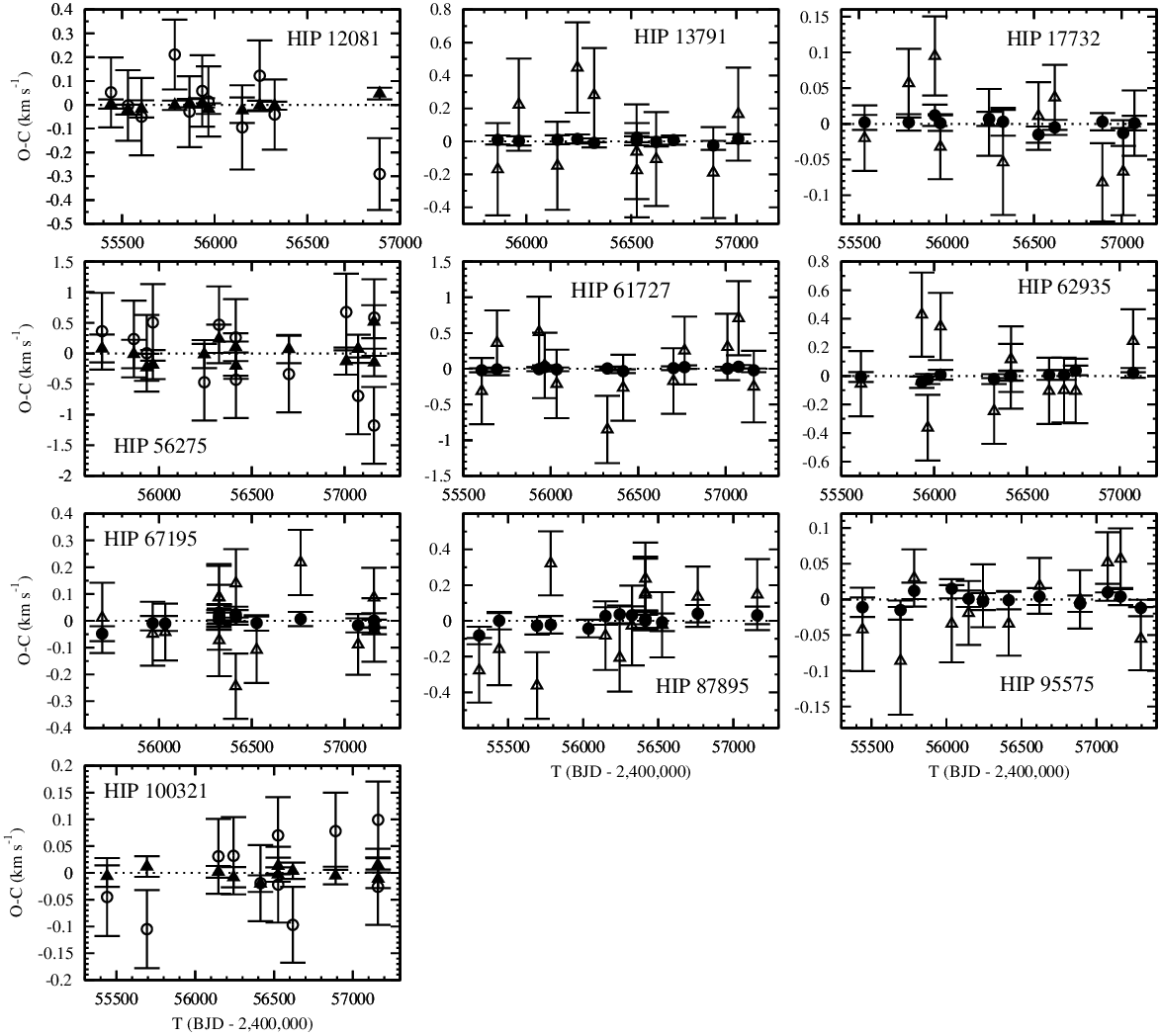} 
\caption{The residuals of the RVs obtained from {\sc todmor} for the 10 SB2s. The circles refer to the primary component, and the triangles to the secondary component. For readability, the residuals of the most accurate RV measurements are in filled symbols.}
\label{fig:resOrbSB2}
\end{figure*}

The orbital solutions, including previously published measurements, are displayed on Fig.\,\ref{fig:orbSB2}. The ($O-C$) residuals are included in Table~\ref{tab:RVs} and are plotted on Fig.\,\ref{fig:resOrbSB2}. We do not observe any drift for any of the 10 SB2.

\section{Combined orbital solution and masses for HIP\,87895}
\label{sec:HIP87895}

\begin{figure}
\includegraphics[clip=,height=5.4 in]{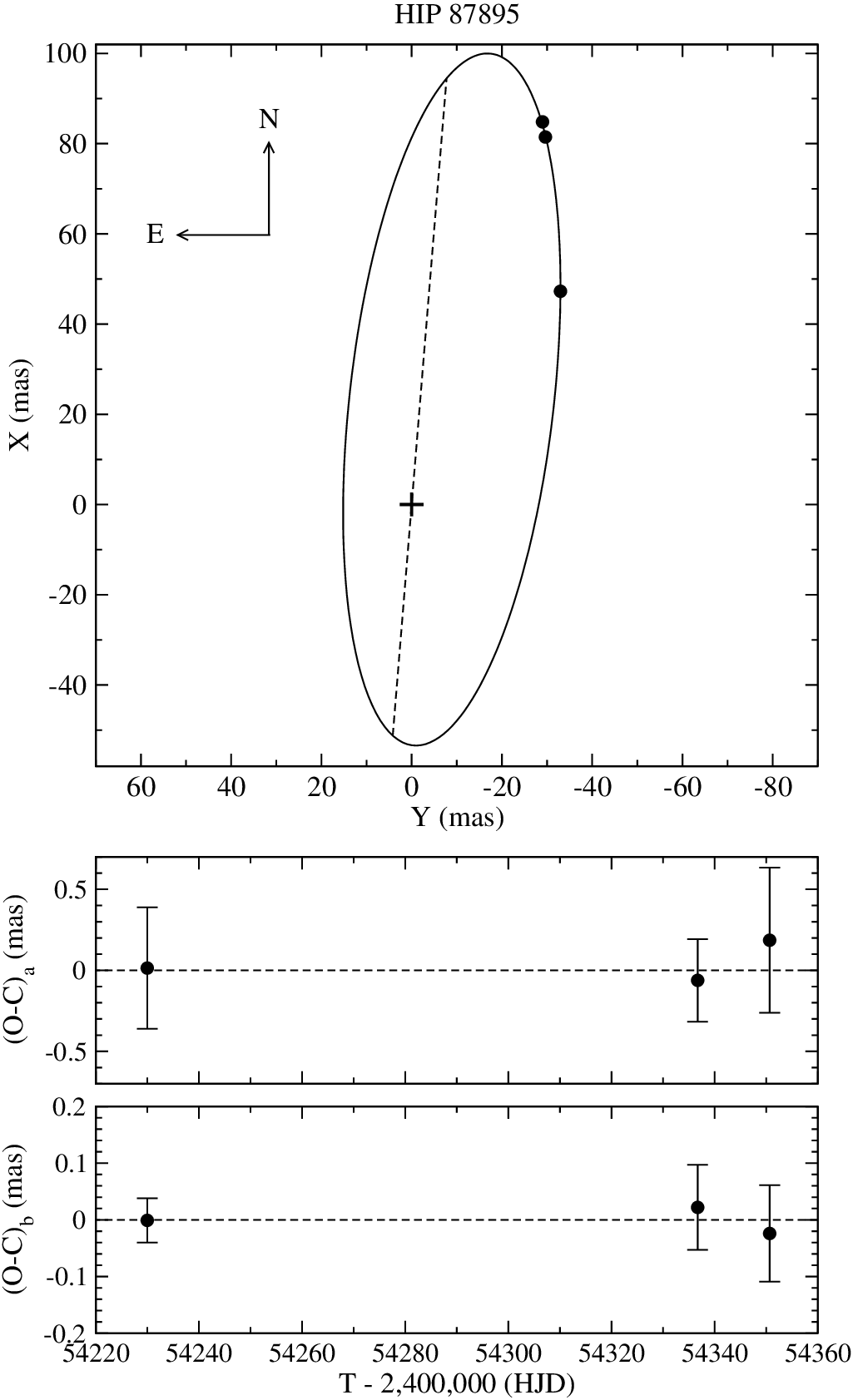}
 \caption{The visual part of the combined orbit of HIP 87895.Upper panel: the
visual orbit; the circles are the 3 positions obtained from long-base interferometry; the node line is in dashes.
Middle panel: the residuals along the semi-major axis of the error ellipsoid. Lower panel: the residuals along the
semi-minor axis of the error ellipsoid.
}
\label{fig:BV-HIP87895}
\end{figure}

\begin{table}
 \caption{The combined VB+SB2 solutions of HIP 87895; For consistency with the SB orbits and with the
forthcoming astrometric orbit, $\omega$ refer to the motion of the primary
component.}
\begin{tabular}{@{}lc@{}}
\hline
  &  HIP 87895 \\
\hline
$P$ (days)              &  881.628 $\pm$  0.064        \\
$T_0$ (BJD-2400000)      & 55650.39 $\pm$ 0.38        \\
$e$                     & 0.4165 $\pm$ 0.0010   \\
$V_0$ (km~s$^{-1}$)            & -32.347 $\pm$ 0.014    \\
$\omega_1$ ($^{\rm o}$) & 135.46 $\pm$ 0.16 \\
$\Omega$($^{\rm o}$; eq. 2000) &  175.32 $\pm$ 0.44 \\
$i$  ($^{\rm o}$)       & 72..83 $\pm$ 0.47 \\ex
$a$ (mas) & 80.64 \\
${\cal M}_1$ (${\cal M}_\odot$) & 1.132 $\pm$ 0.014 \\
${\cal M}_2$ (${\cal M}_\odot$) & 0.7421 $\pm$ 0.0073 \\
$\varpi$ (mas)          & 36.35 $\pm$ 0.20 \\
$d_{n-p}$ (km~s$^{-1}$) & 0.456 $\pm$ 0.047 \\
$d_{2-1\;p}$ (km~s$^{-1}$) & -0.754 $\pm$ 0.608 \\
$d_{2-1\;n}$ (km~s$^{-1}$) & 0.320 $\pm$ 0.056 \\
$\sigma_{(o-c)\;VLTI}$ (mas)  &  0.086 \\
$\sigma_{(o-c)\;RV\;p}$ (km~s$^{-1}$)  & 0.841, 1.211 \\
$\sigma_{(o-c)\;RV\;n}$ (km~s$^{-1}$)  & 0.035, 0.209 \\
\hline
\end{tabular}
\label{tab:HIP87895}
\end{table}

HIP~87895 is the close visual binary (VB) MCA 50. Among the many observations recorded in the on-line {\it Fourth
Catalogue of Interferometric Measurements of Binary Stars}\footnote{http://ad.usno.navy.mil/wds/int4.html}, we
found 3 very accurate long-base interferometric measurements performed with the {\it Palomar Testbed Interferometer} (PTI).
Since the RV measurements are already providing parameters common to the spectroscopic and to the visual orbits,
three 2-dimension measurements are quite sufficient to derive the remaining ones, which are: the orbital
inclination, $i$, the position angle of the nodal line, $\Omega$, and the apparent semi-major axis or the
trigonometric parallax (both may equally be used).

These measurements and the rescaled error ellipsoids provided by \cite{Muterspaugh} are taken into account
simultaneously with our RV measurements, in order to directly derive the masses of the components and the
trigonometric parallax of the system.

The results are given in Table~\ref{tab:HIP87895}, and in Fig.~\ref{fig:BV-HIP87895}. We derived the two masses with accuracies of 1.2 and 0.98~\%, respectively. 

\section{Summary and conclusion}
\label{sec:conclusion}

We obtained for 10 SB2 new radial velocity measurements from spectra taken with SOPHIE, among which four new SB2 identified in paper I. The {\sc todmor} algorithm was used to separate the two components in each spectrum. All ten systems had previously published measurements in archive, which we added to our measurements to calculate the orbital solutions. We also derived estimations of stellar parameters obtained by optimizing the two dimensional cross-correlation obtained with {\sc todmor}. 

We achieved the objective of deriving minimum mass with an accuracy better than 1\% for 15 of the 20 stars of our SB2 sample. Moreover, all ten binaries have the minimum mass of both components estimated with relative uncertainties lower than 1.2\%. This is a great achievement of the combination of {\sc todmor} and SOPHIE. Especially it did very well in extreme configurations such as with small SNR ($\sim$40), and very small flux ratio ($\alpha$$<$$0.05$) between the two components.

We combined our RV measurements of HIP~87895 with 3 relative positions derived from long-based interferometric observations, and we obtained new
mass estimate of the components, $M_1$=$1.132$$\pm $$0.014$\,$M_\odot$ and $M_2$=$0.7421$$\pm $$0.0073$\,$M_\odot$. 
On the basis of speckle observations, \cite{McAlister95} previously derived $M_1$$=$$1.16$$\pm $$0.12$\,$M_\odot$ and $M_2$$=$$0.77$$\pm $$0.05$\,$M_\odot$, i.e. with relative errors of 10 and 6~\%. Therefore, our measurements refine profitably to the 1\% level the confidence range on the masses of HIP\,87895.

Added to the systems observed with {\sc Pionier}, we have now 3 binaries observed with SOPHIE which may be used to check the masses that will be derived from {\sc Gaia}.

\section*{Acknowledgments}

This project was supported by the french INSU-CNRS ``Programme National de Physique Stellaire''
and ``Action Sp\'{e}cifique {\it Gaia}''. We are grateful to the staff of the
Haute--Provence Observatory, and especially to Dr F. Bouchy, Dr H. Le Coroller, Dr M. V\'{e}ron, and the night assistants, for their
kind assistance. We made use of the SIMBAD database, operated at CDS, Strasbourg, France. This research has received funding from the European Community's Seventh Framework Programme (FP7/2007-2013) under grant-agreement numbers 291352 (ERC). The authors are very thankful to the anonymous reviewer for his help to improve the quality of this paper. 

\label{lastpage}

\bsp

\end{document}

%% file: HIP12081_RV_table_latex.tex
\begin{tabular}{@{}l@{~~}c@{~~}c@{~~}c@{~~}c@{~~}c@{~~}c@{}} 
\hline  
\multicolumn{7}{c}{HIP 12081} \\ [1ex]
BJD      & $RV_1$        & $\sigma_{RV 1}$ & $RV_2$        & $\sigma_{RV 2}$ & $O_1-C_1$   & $O_2-C_2$ \\ 
-2400000 & km s$^{-1}$   & km s$^{-1}$      & km s$^{-1}$   & km s$^{-1}$      & km s$^{-1}$ & km s$^{-1}$      \\ 
\hline  
55440.6175    &     -19.3006      &   0.0348            &    4.5327       &   0.0179    & 0.0498 & 0.0052       \\ 
55532.3450    &     5.2930      &   0.0401            &    -22.3033       &   0.0200    & -0.0023 & -0.0206       \\ 
55605.3276    &     -3.1655      &   0.0775            &    -13.1499       &   0.0355    & -0.0492 & -0.0175       \\ 
55784.5923    &     -16.8235      &   0.0334            &    2.0058       &   0.0190    & 0.2120 & -0.0034       \\ 
55864.4742    &     -22.3574      &   0.0424            &    7.7735       &   0.0166    & -0.0294 & 0.0069       \\ 
55933.2302    &     11.2363      &   0.0450            &    -28.6746       &   0.0186    & 0.0576 & 0.0081       \\ 
55966.2854    &     6.7183      &   0.0361            &    -23.8320       &   0.0214    & 0.0144 & -0.0170       \\ 
56148.6082    &     -11.1919      &   0.1043            &    -4.4740       &   0.0530    & -0.0939 & -0.0243       \\ 
56243.3824    &     -18.1414      &   0.0393            &    3.3409       &   0.0191    & 0.1222 & -0.0043       \\ 
56323.3304    &     -20.6003      &   0.0384            &    5.8376       &   0.0160    & -0.0433 & -0.0025       \\ 
56889.6346    &     1.3632      &   0.0490            &    -18.2750       &   0.0237    & -0.2906 & 0.0463       \\ 
&&&&&&   \\ 
&&&&&&   \\ 
&&&&&&   \\ 
\end{tabular}

%% file: HIP13791_RV_table_latex.tex
\begin{tabular}{@{}l@{~~}c@{~~}c@{~~}c@{~~}c@{~~}c@{~~}c@{}} 
\hline  
\multicolumn{7}{c}{HIP 13791} \\ [1ex]
BJD      & $RV_1$        & $\sigma_{RV 1}$ & $RV_2$        & $\sigma_{RV 2}$ & $O_1-C_1$   & $O_2-C_2$ \\ 
-2400000 & km s$^{-1}$   & km s$^{-1}$      & km s$^{-1}$   & km s$^{-1}$      & km s$^{-1}$ & km s$^{-1}$      \\ 
\hline  
55605.3577    &     -21.0198      &   0.0155            &    12.5004       &   0.2343    & -0.0602 & -0.3409       \\ 
55784.6009    &     -8.0152$^*$      &   0.0119$^*$            &    0.2619$^*$       &   0.2294$^*$    & -0.1650$^*$ & 9.5587$^*$       \\ 
55864.4964    &     -0.6383      &   0.0104            &    -21.6297       &   0.1661    & 0.0092 & -0.1697       \\ 
55965.3042    &     4.6405      &   0.0114            &    -30.1618       &   0.1654    & 0.0038 & 0.2217       \\ 
56148.6243    &     -13.0655      &   0.0099            &    -0.6196       &   0.1373    & 0.0105 & -0.1477       \\ 
56243.4109    &     -16.3507      &   0.0099            &    5.5330       &   0.1533    & 0.0154 & 0.4489       \\ 
56323.3367    &     -15.3310      &   0.0114            &    3.6038       &   0.1777    & -0.0089 & 0.2826       \\ 
56525.5659    &     -22.8333      &   0.0117            &    15.9524       &   0.1816    & 0.0041 & -0.0597       \\ 
56526.6076    &     -22.7883      &   0.0113            &    15.7956       &   0.1800    & 0.0231 & -0.1728       \\ 
56618.4809    &     -20.1445      &   0.0113            &    11.3540       &   0.1775    & -0.0036 & -0.1047       \\ 
56701.3400    &     9.2144      &   0.0116            &    -36.7152$^*$       &   0.2370$^*$    & 0.0088 & 1.3839$^*$       \\ 
56890.6072    &     4.3730      &   0.0110            &    -30.1670       &   0.1576    & -0.0229 & -0.1901       \\ 
57009.3890    &     -21.3907      &   0.0112            &    13.7637       &   0.1722    & 0.0158 & 0.1678       \\ 
&&&&&&   \\ 
\end{tabular}

%% file: HIP17732_RV_table_latex.tex
\begin{tabular}{@{}l@{~~}c@{~~}c@{~~}c@{~~}c@{~~}c@{~~}c@{}} 
\hline  
\multicolumn{7}{c}{HIP 17732} \\ [1ex]
BJD      & $RV_1$        & $\sigma_{RV 1}$ & $RV_2$        & $\sigma_{RV 2}$ & $O_1-C_1$   & $O_2-C_2$ \\ 
-2400000 & km s$^{-1}$   & km s$^{-1}$      & km s$^{-1}$   & km s$^{-1}$      & km s$^{-1}$ & km s$^{-1}$      \\ 
\hline  
55532.4636    &     -22.8766      &   0.0095            &    21.9980       &   0.0170    & 0.0019 & -0.0197       \\ 
55784.6076    &     -22.5330      &   0.0105            &    21.6484       &   0.0215    & 0.0016 & 0.0584       \\ 
55933.2851    &     12.7972      &   0.0135            &    -22.2329       &   0.0328    & 0.0123 & 0.0946       \\ 
55965.3166    &     12.1605      &   0.0099            &    -21.5826       &   0.0163    & 0.0003 & -0.0318       \\ 
56243.4535    &     12.1444      &   0.0089            &    -21.5210       &   0.0189    & 0.0060 & 0.0026       \\ 
56323.3438    &     -16.6398      &   0.0178            &    14.2097       &   0.0550    & 0.0020 & -0.0529       \\ 
56525.5908    &     11.5734      &   0.0103            &    -20.8293       &   0.0201    & -0.0154 & 0.0110       \\ 
56618.4465    &     -22.5381      &   0.0096            &    21.6251       &   0.0163    & -0.0060 & 0.0382       \\ 
56890.6164    &     -20.9656      &   0.0110            &    19.5609       &   0.0322    & 0.0023 & -0.0809       \\ 
57009.3996    &     7.0245      &   0.0163            &    -15.2479       &   0.0403    & -0.0123 & -0.0676       \\ 
57073.3157    &     12.5569      &   0.0093            &    -22.0421       &   0.0166    & 0.0009 & 0.0008       \\ 
&&&&&&   \\ 
&&&&&&   \\ 
&&&&&&   \\ 
\end{tabular}

%% file: HIP56275_RV_table_latex.tex
\begin{tabular}{@{}l@{~~}c@{~~}c@{~~}c@{~~}c@{~~}c@{~~}c@{}} 
\hline  
\multicolumn{7}{c}{HIP 56275} \\ [1ex]
BJD      & $RV_1$        & $\sigma_{RV 1}$ & $RV_2$        & $\sigma_{RV 2}$ & $O_1-C_1$   & $O_2-C_2$ \\ 
-2400000 & km s$^{-1}$   & km s$^{-1}$      & km s$^{-1}$   & km s$^{-1}$      & km s$^{-1}$ & km s$^{-1}$      \\ 
\hline  
55692.3811    &     7.9257      &   0.0226            &    -64.5112       &   0.1000    & 0.3626 & 0.0800       \\ 
55863.6796    &     -46.4777      &   0.0309            &    35.6378       &   0.1108    & 0.2361 & -0.0068       \\ 
55933.6492    &     6.6220      &   0.0266            &    -63.0758       &   0.0791    & 0.0026 & -0.2274       \\ 
55966.5170    &     -36.2910      &   0.0297            &    17.1534       &   0.1229    & 0.5121 & -0.1887       \\ 
56243.6608    &     -41.0739      &   0.0280            &    24.3508       &   0.1038    & -0.4721 & -0.0067       \\ 
56323.5683    &     -2.5587      &   0.0224            &    -44.7952       &   0.1061    & 0.4674 & 0.2403       \\ 
56413.3606    &     6.2006      &   0.0221            &    -61.4924       &   0.0961    & 0.2601 & 0.1022       \\ 
56414.3535    &     4.4350      &   0.0239            &    -59.8103       &   0.0548    & -0.4309 & -0.2002       \\ 
56700.5640    &     5.7582      &   0.0178            &    -61.8093       &   0.1047    & -0.3363 & 0.0697       \\ 
57009.6387    &     -39.1607      &   0.0248            &    22.8202       &   0.0789    & 0.6746 & -0.1217       \\ 
57073.4672    &     -7.8378      &   0.0291            &    -37.3543       &   0.1066    & -0.6925 & 0.0742       \\ 
57159.4023    &     -49.4717      &   0.0484            &    39.0880       &   0.1873    & -1.1735 & 0.5173       \\ 
57160.3715    &     -46.9715      &   0.0302            &    37.0594       &   0.0909    & 0.5892 & -0.1492       \\ 
&&&&&&   \\ 
\end{tabular}

%% file: HIP61727_RV_table_latex.tex
\begin{tabular}{@{}l@{~~}c@{~~}c@{~~}c@{~~}c@{~~}c@{~~}c@{}} 
\hline  
\multicolumn{7}{c}{HIP 61727} \\ [1ex]
BJD      & $RV_1$        & $\sigma_{RV 1}$ & $RV_2$        & $\sigma_{RV 2}$ & $O_1-C_1$   & $O_2-C_2$ \\ 
-2400000 & km s$^{-1}$   & km s$^{-1}$      & km s$^{-1}$   & km s$^{-1}$      & km s$^{-1}$ & km s$^{-1}$      \\ 
\hline  
55605.5848    &     -12.6740      &   0.0136            &    15.9258       &   0.1301    & -0.0215 & -0.3118       \\ 
55693.3588    &     6.5327      &   0.0074            &    -13.5889       &   0.0846    & -0.0088 & 0.3627       \\ 
55933.6598    &     -18.4952      &   0.0158            &    25.9363       &   0.2306    & -0.0041 & 0.5154       \\ 
55965.6569    &     10.2366      &   0.0092            &    -19.6786       &   0.1063    & 0.0244 & 0.0466       \\ 
56034.4583    &     -36.6083      &   0.0087            &    53.6952       &   0.1865    & -0.0076 & -0.2096       \\ 
56323.6069    &     3.8442      &   0.0121            &    -10.5547       &   0.1624    & 0.0030 & -0.8501       \\ 
56413.4059    &     -12.8533      &   0.0134            &    16.2369       &   0.1274    & -0.0334 & -0.2641       \\ 
56700.6043    &     -25.6469      &   0.0089            &    36.5198       &   0.1158    & 0.0088 & -0.1701       \\ 
56764.3934    &     7.9839      &   0.0082            &    -15.9285       &   0.1737    & 0.0235 & 0.2550       \\ 
57009.6713    &     8.0070      &   0.0086            &    -15.9493       &   0.1395    & 0.0020 & 0.3044       \\ 
57073.5033    &     -19.5056      &   0.0071            &    27.7723       &   0.2897    & 0.0283 & 0.7111       \\ 
57159.4393    &     18.4660      &   0.0089            &    -32.9927       &   0.2474    & -0.0225 & -0.2502       \\ 
&&&&&&   \\ 
&&&&&&   \\ 
\end{tabular}

%% file: HIP62935_RV_table_latex.tex
\begin{tabular}{@{}l@{~~}c@{~~}c@{~~}c@{~~}c@{~~}c@{~~}c@{}} 
\hline  
\multicolumn{7}{c}{HIP 62935} \\ [1ex]
BJD      & $RV_1$        & $\sigma_{RV 1}$ & $RV_2$        & $\sigma_{RV 2}$ & $O_1-C_1$   & $O_2-C_2$ \\ 
-2400000 & km s$^{-1}$   & km s$^{-1}$      & km s$^{-1}$   & km s$^{-1}$      & km s$^{-1}$ & km s$^{-1}$      \\ 
\hline  
55605.6068    &     -12.3906      &   0.0106            &    19.1954       &   0.1083    & -0.0079 & 0.5511       \\ 
55933.7034    &     12.2587      &   0.0212            &    -16.6646       &   0.2002    & -0.0453 & 1.0353       \\ 
55965.6957    &     10.9649      &   0.0108            &    -15.5176       &   0.1105    & -0.0205 & 0.2413       \\ 
56034.4799    &     -9.0670      &   0.0119            &    14.7234       &   0.1188    & 0.0067 & 0.9506       \\ 
56323.6219    &     -4.1875      &   0.0106            &    6.9054       &   0.1110    & -0.0224 & 0.3592       \\ 
56413.4178    &     -10.6338      &   0.0108            &    16.6760       &   0.1052    & 0.0037 & 0.6010       \\ 
56414.3653    &     -10.9714      &   0.0109            &    17.2867       &   0.1098    & -0.0001 & 0.7203       \\ 
56619.6800    &     6.6032      &   0.0110            &    -8.7968       &   0.1128    & 0.0060 & 0.5014       \\ 
56700.6311    &     -12.9241      &   0.0105            &    19.9558       &   0.1044    & 0.0065 & 0.5048       \\ 
56763.4730    &     9.7161      &   0.0103            &    -13.3344       &   0.1035    & 0.0370 & 0.5012       \\ 
57073.5373    &     14.0341      &   0.0094            &    -19.3651       &   0.0930    & 0.0212 & 0.8509       \\ 
&&&&&&   \\ 
&&&&&&   \\ 
&&&&&&   \\ 
\end{tabular}

%% file: HIP67195_RV_table_latex.tex
\begin{tabular}{@{}l@{~~}c@{~~}c@{~~}c@{~~}c@{~~}c@{~~}c@{}} 
\hline  
\multicolumn{7}{c}{HIP 67195} \\ [1ex]
BJD      & $RV_1$        & $\sigma_{RV 1}$ & $RV_2$        & $\sigma_{RV 2}$ & $O_1-C_1$   & $O_2-C_2$ \\ 
-2400000 & km s$^{-1}$   & km s$^{-1}$      & km s$^{-1}$   & km s$^{-1}$      & km s$^{-1}$ & km s$^{-1}$      \\ 
\hline  
55693.4447    &     -25.3810      &   0.0201            &    17.8880       &   0.1602    & -0.0487 & 0.0120       \\ 
55965.7213    &     -24.1047      &   0.0187            &    15.6821       &   0.1456    & -0.0092 & -0.0474       \\ 
56033.4718    &     -16.6905      &   0.0198            &    2.8186       &   0.1299    & -0.0100 & -0.0424       \\ 
56323.6540    &     -25.6179      &   0.0185            &    18.4821       &   0.1453    & 0.0131 & 0.0876       \\ 
56323.7172    &     -25.6252      &   0.0185            &    18.4805       &   0.1502    & 0.0034 & 0.0902       \\ 
56324.4871    &     -25.4322      &   0.0209            &    18.0313       &   0.1648    & 0.0304 & -0.0709       \\ 
56324.5606    &     -25.4128      &   0.0196            &    18.0605       &   0.1480    & 0.0177 & 0.0141       \\ 
56413.4778    &     3.2056      &   0.0197            &    -31.8507       &   0.1488    & 0.0263 & -0.2456       \\ 
56414.4222    &     0.2645      &   0.0192            &    -26.3836       &   0.1566    & 0.0138 & 0.1391       \\ 
56526.3286    &     55.7540      &   0.0209            &    -122.9667       &   0.1516    & -0.0025 & -0.1150       \\ 
56763.4478    &     26.5703      &   0.0184            &    -71.9675       &   0.1492    & 0.0100 & 0.2148       \\ 
57073.5916    &     -18.4045      &   0.0182            &    5.7351       &   0.1389    & -0.0178 & -0.0869       \\ 
57159.4122    &     4.9344      &   0.0192            &    -34.5608       &   0.1357    & 0.0026 & 0.0858       \\ 
57160.4048    &     1.4943      &   0.0193            &    -28.7538       &   0.1452    & -0.0216 & -0.0354       \\ 
\end{tabular}

%% file: HIP87895_RV_table_latex.tex
\begin{tabular}{@{}l@{~~}c@{~~}c@{~~}c@{~~}c@{~~}c@{~~}c@{}} 
\hline  
\multicolumn{7}{c}{HIP 87895} \\ [1ex]
BJD      & $RV_1$        & $\sigma_{RV 1}$ & $RV_2$        & $\sigma_{RV 2}$ & $O_1-C_1$   & $O_2-C_2$ \\ 
-2400000 & km s$^{-1}$   & km s$^{-1}$      & km s$^{-1}$   & km s$^{-1}$      & km s$^{-1}$ & km s$^{-1}$      \\ 
\hline  
55306.5401    &     -25.5851      &   0.0102            &    -42.7398       &   0.1192    & -0.0824 & -0.2769       \\ 
55440.3398    &     -24.3803      &   0.0107            &    -44.3309       &   0.1369    & 0.0012 & -0.1582       \\ 
55692.5686    &     -47.1465      &   0.0103            &    -9.8598       &   0.1237    & -0.0269 & -0.3621       \\ 
55783.4403    &     -41.7373      &   0.0102            &    -17.4182       &   0.1176    & -0.0231 & 0.3225       \\ 
56034.6403    &     -29.2514      &   0.0104            &    -37.8613$^*$       &   0.1264$^*$    & -0.0441 & -1.0478$^*$       \\ 
56148.4136    &     -26.2422      &   0.0103            &    -41.3782       &   0.1288    & 0.0258 & -0.0824       \\ 
56243.2450    &     -24.6700      &   0.0105            &    -43.8842       &   0.1276    & 0.0358 & -0.2061       \\ 
56324.6564    &     -24.3659      &   0.0121            &    -44.1758       &   0.1547    & 0.0306 & -0.0261       \\ 
56413.6297    &     -26.9424      &   0.0104            &    -40.1199       &   0.1358    & 0.0005 & 0.1468       \\ 
56414.5000    &     -26.9901      &   0.0104            &    -39.9479       &   0.1370    & 0.0080 & 0.2346       \\ 
56414.5337    &     -26.9911      &   0.0104            &    -40.0227       &   0.1364    & 0.0091 & 0.1564       \\ 
56525.3681    &     -42.7839      &   0.0103            &    -16.1425       &   0.1206    & -0.0066 & -0.0228       \\ 
56763.6370    &     -35.3359      &   0.0103            &    -27.2716       &   0.1088    & 0.0394 & 0.1357       \\ 
57159.5261    &     -24.3802      &   0.0103            &    -43.9809       &   0.1341    & 0.0306 & 0.1471       \\ 
\end{tabular}

%% file: HIP95575_RV_table_latex.tex
\begin{tabular}{@{}l@{~~}c@{~~}c@{~~}c@{~~}c@{~~}c@{~~}c@{}} 
\hline  
\multicolumn{7}{c}{HIP 95575} \\ [1ex]
BJD      & $RV_1$        & $\sigma_{RV 1}$ & $RV_2$        & $\sigma_{RV 2}$ & $O_1-C_1$   & $O_2-C_2$ \\ 
-2400000 & km s$^{-1}$   & km s$^{-1}$      & km s$^{-1}$   & km s$^{-1}$      & km s$^{-1}$ & km s$^{-1}$      \\ 
\hline  
55306.6052    &     -78.1722$^*$      &   0.0068$^*$            &    -50.6807$^*$       &   0.0274$^*$    & -0.9557$^*$ & -0.9533$^*$       \\ 
55440.3855    &     -69.9864      &   0.0107            &    -57.8756       &   0.0514    & -0.0113 & -0.0413       \\ 
55693.5658    &     -60.6375      &   0.0103            &    -68.3909       &   0.0740    & -0.0148 & -0.0863       \\ 
55784.4181    &     -74.6558      &   0.0069            &    -52.5504       &   0.0199    & 0.0121 & 0.0302       \\ 
56034.5983    &     -57.5141      &   0.0103            &    -71.8023       &   0.0453    & 0.0156 & -0.0350       \\ 
56147.4329    &     -76.3749      &   0.0070            &    -50.6872       &   0.0298    & 0.0013 & -0.0191       \\ 
56243.2640    &     -51.3445      &   0.0075            &    -78.6904       &   0.0286    & -0.0033 & 0.0051       \\ 
56414.5737    &     -53.1692      &   0.0091            &    -76.6845       &   0.0299    & -0.0014 & -0.0339       \\ 
56619.2736    &     -74.8908      &   0.0075            &    -52.3081       &   0.0177    & 0.0033 & 0.0192       \\ 
56890.4483    &     -50.3115      &   0.0067            &    -79.8551       &   0.0221    & -0.0060 & -0.0002       \\ 
57073.6986    &     -50.2649      &   0.0077            &    -79.8382       &   0.0248    & 0.0093 & 0.0518       \\ 
57159.5432    &     -73.8810      &   0.0079            &    -53.3993       &   0.0257    & 0.0044 & 0.0573       \\ 
57295.3098    &     -76.9201      &   0.0069            &    -50.1277       &   0.0290    & -0.0122 & -0.0548       \\ 
&&&&&&   \\ 
\end{tabular}

%% file: HIP100321_RV_table_latex.tex
\begin{tabular}{@{}l@{~~}c@{~~}c@{~~}c@{~~}c@{~~}c@{~~}c@{}} 
\hline  
\multicolumn{7}{c}{HIP 100321} \\ [1ex]
BJD      & $RV_1$        & $\sigma_{RV 1}$ & $RV_2$        & $\sigma_{RV 2}$ & $O_1-C_1$   & $O_2-C_2$ \\ 
-2400000 & km s$^{-1}$   & km s$^{-1}$      & km s$^{-1}$   & km s$^{-1}$      & km s$^{-1}$ & km s$^{-1}$      \\ 
\hline  
55440.4025    &     17.4467      &   0.0262            &    -20.6978       &   0.0354    & -0.0453 & -0.0053       \\ 
55693.6211    &     -30.5121      &   0.0269            &    42.2770       &   0.0338    & -0.1057 & 0.0121       \\ 
56147.4444    &     -33.9387      &   0.0172            &    46.9496       &   0.0192    & 0.0303 & 0.0021       \\ 
56243.2764    &     32.2481      &   0.0242            &    -40.0539       &   0.0329    & 0.0321 & -0.0083       \\ 
56414.6090    &     -30.0393      &   0.0214            &    41.7375       &   0.0272    & -0.0196 & -0.0191       \\ 
56525.4060    &     -35.5108      &   0.0223            &    49.0625       &   0.0234    & 0.0698 & -0.0032       \\ 
56526.4184    &     -34.6986      &   0.0204            &    47.8903       &   0.0271    & -0.0222 & 0.0130       \\ 
56619.4231    &     26.0449      &   0.0204            &    -32.0577       &   0.0268    & -0.0975 & 0.0049       \\ 
56890.5147    &     32.8145      &   0.0242            &    -40.7347       &   0.0288    & 0.0780 & -0.0049       \\ 
57159.5852    &     23.8836      &   0.0215            &    -29.1134       &   0.0249    & -0.0255 & 0.0137       \\ 
57160.5816    &     18.7741      &   0.0229            &    -22.2593       &   0.0302    & 0.0997 & -0.0127       \\ 
&&&&&&   \\ 
&&&&&&   \\ 
&&&&&&   \\ 
\end{tabular}